# Automated Root Cause Analysis System for Complex Data Products

Mathieu Demarne, Miso Cilimdzic, Tom Falkowski, Timothy Johnson, Jim Gramling, Wei Kuang, Hoobie Hou, Amjad Aryan, Gayatri Subramaniam, Kenny Lee, Manuel Mejia, Lisa Liu, Divya Vermareddy

Microsoft
One Microsoft Way
Redmond, WA 98052 USA
{mdemarne, misoc, tomfalko, tijohhnso, jigramli, weikuang, hsinyuhou, amjadaryan, gsubramaniam, leekenny, manuelmejia, lisaliu, dvermareddy}@microsoft.com

## ABSTRACT

We present ARCAS (Automated Root Cause Analysis System), a diagnostic platform based on a Domain Specific Language (DSL) built for fast diagnostic implementation and low learning curve. Arcas is composed of a constellation of automated troubleshooting guides (Auto-TSGs) that can execute in parallel to detect issues using product telemetry and apply mitigation in near-real-time. The DSL is tailored specifically to ensure that subject matter experts can deliver highly curated and relevant Auto-TSGs in a short time without having to understand how they will interact with the rest of the diagnostic platform, thus reducing time-to-mitigate and saving crucial engineering cycles when they matter most. This contrasts with platforms like Datadog and New Relic, which primarily focus on monitoring and require manual intervention for mitigation. ARCAS uses a Large Language Model (LLM) to prioritize Auto-TSGs outputs and take appropriate actions, thus suppressing the costly requirement of understanding the general behavior of the system. We explain the key concepts behind ARCAS and demonstrate how it has been successfully used for multiple products across Azure Synapse Analytics and Microsoft Fabric Synapse Data Warehouse.

## 1  INTRODUCTION

Complex cloud-based data systems have an ever-greater surface area due to the multiplicity of their components, distributed architecture, and increasing complexity. Diagnosing and troubleshooting such a system is time-consuming, but it is also time-sensitive when the system is not behaving as expected. Site Reliability Engineers (SREs) typically require extensive knowledge of the data system architecture, troubleshooting guides (TSGs), and the history of system failures. For new and senior SREs alike, building and maintaining this knowledge is non-trivial as it requires sustaining a global understanding of how the product is evolving, of the relationship with customers, and of past issues. As a result, many incidents take longer to resolve than they ought to, driving customer satisfaction down, and with it, growing engineering frustration.

However, automating such investigative work through diagnostics and proactive monitoring is challenging. The interoperability of the multiple layers of the product stack means that engineers can have siloed knowledge of components even if those components influence each other. To make matters more complex, cloud-native applications are typically backed by a variety of products, each of which can impact performance and availability. This means that incidents can be misassigned and need to manually be routed to other teams. A study across multiple Microsoft services by J. Chen et al. show that many incidents are re-assigned, which increases the time to mitigate them **[1]**.

Rule-based pattern-matching algorithms such as Rete **[2]** have been used in the industry to build automated diagnostic systems. Similarly, interactive, web-based data dashboards are common to display the internal behavior of large systems to allow SREs to quickly understand their current state. Moreover, those automated diagnostic and reporting systems are often coupled with monitors producing alerts when the system starts to misbehave – some of which require manual intervention.

Often, those supporting systems rely on different architectures. Automated diagnostics are typically built on rule-based engines, dashboards typically rely on web interfaces, and monitors and watchdogs depend on metrics and custom code logic to interact with the product in production. With little overlap, those components make the work of SREs and subject matter experts more complicated as they need to understand software development lifecycles (SDLCs) for multiple distinct internal support products. In addition, some piece of automation logic can typically be used across all these components, causing additional duplication as those tend to be siloed. These platforms typically stand at the intersection of site reliability engineering and product engineering as they must encapsulate the knowledge from subject matter experts but are typically decoupled from the main product code.

In this paper, we present a new diagnostic platform unifying automated diagnostics, data rendering, and monitoring into a single domain-specific language. Our system does not require a non-initiated engineer to understand existing diagnostics to implement new ones, thus saving crucial engineering time and reducing the cost of implementation and learning curve. The interaction between a newly created diagnostic and existing Auto-TSGs is mostly done under-the-hood.

In contrast to the traditional approach of using separate tools for various aspects of system management, ARCAS offers a unified solution, distinguishing itself from platforms like Datadog [3] and New Relic [4] that are primarily used for monitoring and troubleshooting but lack an integrated system for automated mitigation and step-by-step diagnostic automation.

We show how useful our platform has been for Azure Synapse Analytics [5] and, more recently, Microsoft Fabric Synapse Data Warehouse [6]. We also explain how our system unifies and integrates with existing support portals for both product group engineers and customer service and support engineers.

This paper is organized as follows. *Section 2* covers the motivation behind our platform. *Section 3* covers its design and intuition. *Section 4* covers the implementation details of our Auto-TSGs, focusing on our DSL language. *Section 5* explains how this solution has been successfully used within Microsoft products. *Section 6* covers related work. Finally, *section 7* covers future effort and *section 8* concludes.

## 2  MOTIVATION

Cloud-native complex data systems typically rely on an array of sub-products that all need to be up and running for the system to perform well. For an analytical database, such sub-systems might be data storage, the network connectivity among cluster nodes, the gateway-specific stack, the microservice infrastructure on which the system is deployed, the control plane infrastructure that orchestrates provisioning, the platform performing operating system maintenance, and, finally, the billing layer. Each sub-system comes with its own set of challenges, development lifecycle, and team ownership. Some of them are internal only, others, for instance storage, might be available directly to customers through a separate public offering. In addition to this, large data systems are typically distributed, spanning multiple nodes and components, and have their own built-in intelligence to manage resource utilization and allocation. Such systems heavily rely on telemetry (metrics and logs), available in near-real-time. Often, telemetry is specific to one of those sub-layers or components and uses a distinct format adapted to it.

Such complex systems are hard to support without the right documentation. To make matters worse, subject matter experts typically need to have a deep understanding of how components work together to develop new troubleshooting guides, understand how those guides fit in the landscape of other related documents, and make sure that SREs know where to look to find the right information lineage to diagnose and resolve problems.

To help identify issues, SREs typically rely on three main platforms using telemetry as a source:

1. **Dashboards**, typically UI-based pages displaying an array of metrics and logs and optimized for easy navigation between telemetry of related components.

2. **Automated diagnostic systems**, which often are rule-based engines tailored to detect one or more problems from component telemetry. Such systems are typically called independently or integrated into dashboards and focus on a subset of aspects.

3. **Monitors** that proactively read telemetry and detect issues faster than customers, create incidents or, in some cases, apply automated mitigations.

The design requirements for each of these systems are typically not the same. Dashboards have interactive UIs and are thus built on web frameworks and libraries (such as React [7], or Node.JS [8]) when not provided by a third-party tool like Prometheus [9] or Grafana [10] or different SaaS solutions; automated systems are typically rule-based; and monitors have near-real-time requirements as well as a need to run in secure environments to operate on production systems directly.

This means that subject matter experts need to:

a. Understand how the component for which they are experts interoperates with other components.

b. Understand how to develop on all three distinct support platforms mentioned above.

c. Understand where and how to integrate new pieces of telemetry logic into the current landscape of those systems to form a coherent whole that non-expert engineers can utilize when on call.

d. Deal with duplication of work when the same piece of telemetry logic is used in dashboards, rule-based diagnostics, and automation.

This can be a significant barrier to overcome, especially when working on time sensitive issues. Such situations can push engineers to bypass implementing diagnostics altogether and instead rely on text notes, which are not always properly contextualized. This, in turn, can increase the time that non-expert SREs spend working on related problems and does not eliminate repetitive, manual work that could otherwise be automated (in the context of site reliability engineering, this is known as toil).

The automated root cause analysis system presented in this paper aims to tackle each and every one of those challenges, by providing a single unifying domain-specific language (DSL) that is easy to learn and that can provide both rich data rendering (for (**1**)), complex diagnostic automations (for (**2**)), the ability to run on a schedule (for (**3**)) and perform automated actions (for both (**2**) and (**3**)). With a global understanding of a system's architecture, ARCAS can intelligently understand how different pieces of diagnostics need to interoperate, thus removing the need for subject matter experts to understand how the component they own fits with the entire system (for **(a)** and **(c)**) and provide all of it in a single tool (for **(b)** and **(d)**).

# 3 DESIGN

This section covers the design principles behind ARCAS and how they help solve some of the challenges above. We also delve into how engineers typically model troubleshooting guides, and what it means for the DSL language design we decided to implement.

## 3.1 Principles

Through the development of ARCAS, we set some principles that the system had to follow for effectiveness, user-friendliness, and maintainability.

*3.1.1 Low Learning Curve.* ARCAS is centered on maximizing engineering productivity by making Auto-TSGs easy to understand and implement. This principle ensures that engineers, regardless of their familiarity with the system, can quickly learn to read, write, and modify Auto-TSGs. The goal is to reduce the time and effort required to get up to speed with ARCAS, thereby enabling participation from a broader range of contributors in the troubleshooting process. This, in turn, allows for faster implementation for an automated diagnostic when the issue is time-sensitive, such as when a new regression is shipped as part of a new deployment, which reduces toil.

*3.1.2 Standalone.* Each Auto-TSG should be self-contained and comprehensible without the need to grasp the complete collection it is a part of. This approach allows for easier maintenance and updates, as engineers can focus on an individual Auto-TSG without the risk of unintended side effects on other parts of the system. It promotes modularity and simplifies the process of troubleshooting by providing clear and concise instructions within each guide.

*3.1.3 Integrable.* Auto-TSGs need to work together and share context in a standardized format. This interoperability is essential for creating a cohesive troubleshooting experience where different Auto-TSGs can call upon each other as needed. By adhering to this principle, ARCAS can leverage the collective intelligence of multiple TSGs, allowing for more comprehensive diagnostics and solutions.

*3.1.4 Meaningful.* Auto-TSGs should not only return data but also provide textual context that helps non-experts understand the steps taken during the troubleshooting process. This context can be tailored to product engineers, customer support, or even customers depending on a specific audience. Adding value to the raw data by interpreting it and presenting it in a way that is informative and actionable ensures that the output of Auto-TSGs is not just a collection of data points but a narrative that guides users towards understanding and resolving the issue.

*3.1.5 Reproducible.* Finally, an Auto-TSG needs to be repeatable. This principle prioritizes the use of telemetry-based queries as first-class citizens, ensuring that they are documented and can be executed again with the same results (the queries thus need to always be deterministic). Reproducibility is key to verifying findings, sharing knowledge, and ensuring consistency across different troubleshooting sessions.

## 3.2 Modelling Troubleshooting Guides

An Auto-TSG within ARCAS is modeled as a directed acyclic graph to streamline the diagnostic process. We chose this structure because it mirrors the logical flow of troubleshooting (Illustrated in *figure 1*), where each step or node in the graph represents a specific check or diagnostic action. The decision graph format allows for a clear visualization of the troubleshooting path, with branches leading to different outcomes based on the results of each check. Often, an Auto-TSG is a decision tree as opposed to a full graph. See *figure 2* for an example of such a decision tree.

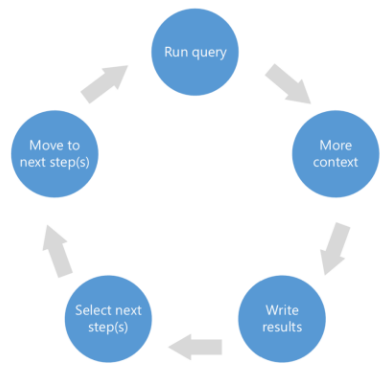

*Figure 1 Typical auto-TSG walkthrough.*

The design of Auto-TSGs as decision graphs enables a context-driven approach to troubleshooting. As the graph is traversed, the context is enriched with data from each node, allowing subsequent decisions to be made with an increasing amount of information. The path taken through the graph is directly influenced by the results of each diagnostic check.

# 4 IMPLEMENTATION OVERVIEW

This section covers the various concepts behind ARCAS. For each section, we explain how they help match the principles presented earlier in this paper.

## 4.1 Configuration-First Domain Specific Language (DSL)

Our domain-specific language is built on top of a YAML-based structure **[11]**. This uses a construction syntax typically familiar to engineers but also abstracts away most of the chaining logic required to execute decision graphs as each step in the graph is one YAML block in a list. Each component has a name, a telemetry query to execute if needed (composed of a source, an explanation, and a query text as well as information required to retain additional context from the retrieved data), and a set of next steps to perform if the query does find matching facts. This abstraction removes the need to define conditional logic blocks altogether, as those are implicitly handled by the framework.

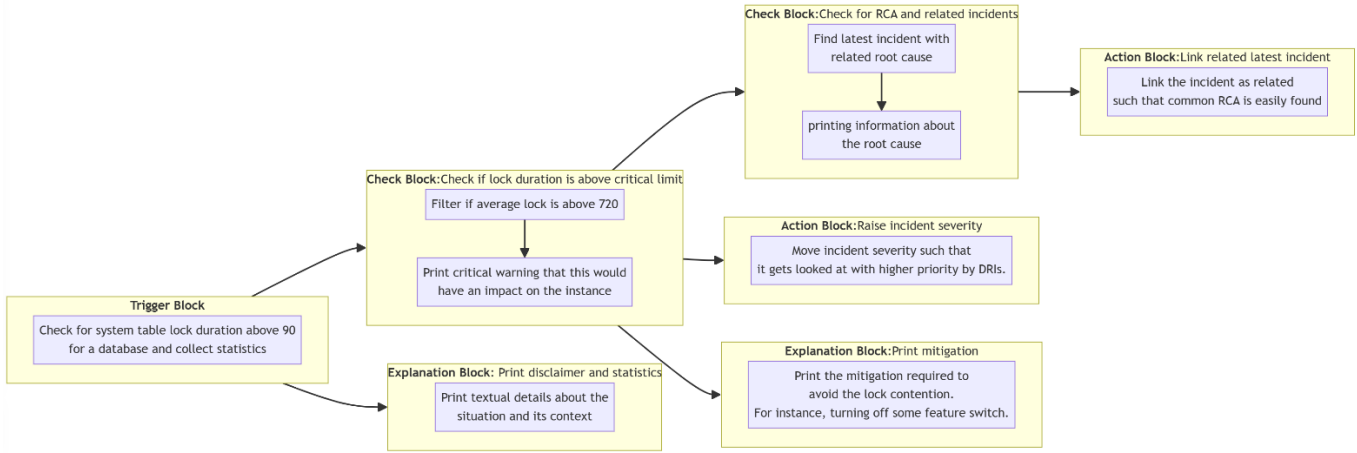

*Figure 2 An investigation tree representing a typical Auto-TSG, in this case related to high lock contention on an Azure Synapse database.*

*Snippet 1* illustrates a simplified Auto-TSG definition for a database system (thus using two context variables, `DatabaseName` and `ServerName` alongside timestamp variables). Please refer to *section 4.2* for an in-depth explanation of its building blocks. The Auto-TSG uses the Kusto Query Language [12] to query telemetry.

```
#########################################################
# UPGRADE INVESTIGATION AUTO-TSG
#########################################################
Metadata:
  Title: Recent Upgrades
  Description: Detects possible recent upgrades.
  Owner: controlplane@someservice.platform.com
  Type: Warning
#########################################################
Triggers:
  - Audiences:
    - InternalTicket
    - InternalOnDemand
    Queries:
      - Source: Kusto
        Explanation: |
          We detected an upgrade for database:
          - Server **{ServerName}**.
          - DB: **{DatabaseName}**
          Here are details of the recent upgrade(s):
        QueryText: |
          ManagementOperations
          | where OperationName == "Upgrade"
          | where ServerName == "{ServerName}"
          | where DatabaseName == "{DatabaseName}"
          | where TimeStamp >= datetime({StartTime})
          | where TimeStamp <= datetime({TimeStamp})
          | summarize UpgradeStart = min(TimeStamp),
                      UpgradeEnd   = max(TimeStamp),
                      State = arg_max(TimeStamp, State)
                   by OperationId
          | extends Duration = UpgradeEnd - UpgradeStart
        AddedContext:
          OperationId: long
          Duration: timespan
          State: string
    NextSteps:
      - check-version-change
      - print-warning-if-long-duration-and-running
#########################################################
Checks:
  - Name: check-version-change
    Query:
      Source: Kusto
      Explanation: >-
        The instance changed versions:
      QueryText: |
        RawDatabaseLogs
        | where ServerName == "{ServerName}"
        | where DatabaseName == "{DatabaseName}"
        | where TimeStamp >= datetime({StartTime})
        | where TimeStamp <= datetime({TimeStamp})
        | summarize by Version
        | summarize count(), make_list(Version)
        | where count_ > 1
#########################################################
Explanations:
  - Name: print-warning-if-long-duration-and-running
    Filter: >-
      {Duration} > 1h and {State} != "Complete"
    Explanation: >-
      There has been a recent upgrade that:
      - Lasted more than **one hour**
      - Did not complete during that period
      This means the system might not be available.
      We need to investigate with high priority.
    NextSteps:
      - raise-severity
#########################################################
Actions:
  - Name: raise-severity
    Action: IncreaseSeverity
    NewSeverity: A
```

*Snippet 1 Illustration of a basic Auto-TSG.*

In the example above, the Auto-TSG defines a trigger that runs on-demand (through a UI portal) as well as when a new support ticket is created (in which case the details of the automated investigation are added to the ticket itself). The first telemetry query checks for recent upgrades in the given incident period, and if one is found, displays it for awareness. It then checks the new versions and if one or more of the current upgrade operations is taking more than one hour. It does this using additional context extracted from the trigger. If so, it raises the incident severity to make sure the incident is looked at earlier rather than later. This step is implicitly specific to the audience type (on support tickets) and not executed when the Auto-TSG is running for the UI-based troubleshooter.

Apart from some necessary structure, the YAML above shines by focusing on three things: (**1**) explanation texts, using markdown; (**2**) telemetry queries; (**3**) context enhancements and step chaining. The actual code doing any type of branching or context storing is entirely abstracted away. Since the structure is quite simple, it can be learned quickly.

## 4.2 Steps and Chaining Logic

The steps in an Auto-TSG can be of multiple types:

*4.2.1 Triggers.* Triggers define entry points for an Auto-TSG. One Auto-TSG can have multiple triggers, and when a trigger is executed and does find data, the TSG becomes active and subsequent steps are executed.

*4.2.2 Checks.* Checks can run additional queries and expand the context that we know about the current situation. That context can then be passed on to a set of next steps (which can in turn be either a check, an explanation block, or an action). As for every type of step running a telemetry query, the data returned can be richly rendered (as a table or a graph) and decorated with explanation text.

*4.2.3 Explanations.* Those blocks are the simplest of all. Instead of running a query, they print additional text data in the output that will be visible to the user. Those explanation text blocks (like any other throughout the step types) can use the information that was extracted by past steps and triggers to contextualize their output.

*4.2.4 Actions.* Those blocks define actions that can be taken by the framework. Those actions are context dependent. Some can act on production clusters, for instance rebooting processes or creating process image dumps for further investigation. Some others can act on existing support tickets that are being enhanced with the ARCAS output. For instance, if an Auto-TSG finds crucial diagnostic information, it can route a ticket to the appropriate on call team for support or increase the ticket severity to expedite a time-sensitive situation, like in the example above. Some special action also allows calling Auto-TSGs using finding from others.

Each one of these types can also have an additional *filter predicate* that, given the current known context, can let the system know if the step needs to be executed or not. This allows **backward-chaining** (where the step itself defines if it needs to be executed given the context) in addition to **forward-chaining** (where the next steps to execute are clearly written), thus adding more flexibility for the author.

Note that this differs from the Rete algorithm, for which chaining is implicit based on known context (i.e., each iteration in a Rete execution checks which rules can be applied based on the known facts to find new facts). Implicit chaining works well for a single, cohesive set of facts that work on the same context and respect strict naming convention, but is not applicable, nor easy to manage and read, when there are many diagnostic authors and a large surface area to cover.

## 4.3 Context Passing and Expansion

This chapter formalizes how the system behaves when executing steps given a context. We can represent the context as a set of key-value pairs:

$$C = \{(k_1, v_1), (k_2, v_2), \ldots\}$$

At each step, a specific telemetry query is run returning tabular data (which can be represented as a table or a graph in the final output). For each row, we extract additional context (if specified that we need to). Every context enhancement step can thus be defined as a function:

$$step_i(C) = C \to \{C' | C' = C \cup c, c \in query_i(C)\}$$

Where $query_i(C)$ represents the set of values retained and returned by the query ran in the step $i$ (as specified in the `AddedContext` block as in *snippet 1* above). Then, subsequent steps are executed for any *variation* of that context in the returned set, such that:

1. In case of multiple rows containing distinct data, we will execute the next step(s) once per distinct variation.

2. In case of duplicate values among rows, we will execute the step only once (since executing it multiple times with the same context would yield the same data, as we assume every telemetry query to be deterministic).

To do this, we define a notion of *equivalence* among steps. We first define the subset $req_{step_i}(C) = \{(k, v) \mid (k, v) \in C \land k \in req_{step_i}\}$ where $req_{step_i}$ represents the context keys *required* to execute the step $i$, which are inferred from the query, the explanation block, and the filter predicate. We can now define a notion of equivalence:

$$step_i(C_x) \equiv step_i(C_y) \text{ if } req_{step_i}(C_x) = req_{step_i}(C_y)$$

In other words, if two steps execute the same query or render the same data, and use the same values from two distinct contexts, they are equivalent, and the framework will ensure only one is executed. This has two benefits:

1. It simplifies the output, as we do not need to display the same information more than once. This is particularly useful when multiple branches in a diagnostic graph can lead to the same sub-step. If the subsequent steps from two equivalent steps use distinct context values, however, we will propagate the results of the execution twice to make sure the graph is properly traversed.

2. It lowers the pressure on the telemetry engine as it avoids re-running queries when not needed.

## 4.4 Base Context

For an Auto-TSG to start executing, we need to have some base set of known facts about an issue. This base set is typically specific to a product and known by Auto-TSG authors. For instance, if the issue is about a database, we would expect the database and the server's names as well as the incident period to be in that base context, ready to be used in Auto-TSG triggers. For other types of products, the base context we might have at our disposal might be different. For Microsoft Fabric Synapse Data Warehouse, for example, we often speak of workspaces.

Typically, this base context is extracted from an incoming incident. It can also be specified manually when Auto-TSGs are triggered interactively. It is then passed along to every Auto-TSG that applies. Note that if we lack the context required for a given Auto-TSG, it will be skipped. Each Auto-TSG can then execute independently and grow what we know of the situation, print details of their investigations, and finally reach a conclusion. We will cover how the distinct Auto-TSG outputs are merged later in this paper.

## 4.5 Audience Type and Support Platforms

Every step in an Auto-TSG can define one or more specific audience tags for which to execute. This tag is typically linked to a calling environment. For instance, a UI-based portal would rely on a different tag compared to a system automatically adding diagnostics to an incoming customer ticket.

This section lists those audience tags in the context of Azure Synapse Analytics.

*4.5.1 Customer Visible.* The customer-visible audience tag allows to display some parts of Auto-TSGs directly to customers when they are trying to diagnose potential issues. Typically, this type of audience provides only highly curated texts that link to public documentation and does not display internal telemetry, even if this one is used for determination. This can help massively with deflection and customer self-help.

*4.5.2 Internal On-Demand.* The on-demand audience is typically used to display Auto-TSG outputs in an internal UI-based portal. Diagnostics are triggered manually by both support and product engineers alike, and typically display telemetry results and a detailed analysis on the product internals. Insights are grouped by their topic of focus.

*4.5.3 Support Ticket.* When customers create tickets, those are typically routed to customer service and support engineers. Our framework uses the ticket content to run a large battery of diagnostics that the support engineers can have readily available to help the customer, sometimes with customer-ready content that can be shared directly. Some actions can also re-adjust the ticket severity depending on the findings, among others.

*4.5.4 Internal Ticket.* When the product team gets involved, the diagnostics can often refer to more technical product details. ARCAS can also act directly on the ticket to route it to the appropriate product team depending on the findings, ensuring that the right subject matter experts treat tickets aligned to their expertise. Known repair tracking items and known root cause analyses (RCAs) can also be automatically linked to those tickets.

*4.5.5 Scheduled.* Finally, some Auto-TSGs can define triggers that run on a specific schedule, using a slightly enhanced syntax. These triggers can create live site incidents as well as directly take actions on top of the production environment to proactively mitigate problems before they are reported by customers. The next section covers this type of Auto-TSGs in more detail.

All the audiences above can be specified at different nodes within the same Auto-TSG. Triggers do require them explicitly, while checks can derive them from triggers implicitly if not specified. This means that the same piece of logic can serve many distinct types of diagnostic environments, avoiding duplication and accompanying customers, customer support engineers, and product engineers throughout issue resolution.

Moreover, the display policy for each audience typically differs. For example, internal tickets might be populated with links to telemetry queries directly alongside the data retrieved by Auto-TSGs. This, however, would not be the case for customer-facing insights, as customers do not have access nor need to be exposed to internal telemetry.

## 4.6 Auto-TSG Chaining and Prioritization

Auto-TSGs are executed in parallel. We attempt to determine whether their triggers apply, provided the given base context has the right information. If they do, an Auto-TSG is considered "active" and its graph is traversed, generating logs and actions to take, until a termination point is reached. Once all Auto-TSGs have terminated, the ARCAS framework has a full picture of all potential issues related to a given base context, and now needs to understand which findings to prioritize, and which actions, if any, to take or to skip.

This is typically done using a combination of Auto-TSG types alongside an understanding of the topics of a ticket or situation. To return to our database example, an Auto-TSG checking for missed backups is important, but if the ticket is about a degradation of query performance, missed backups are unlikely to be the root cause of the problem. Therefore, prioritizing actions to ensure successful backups, while not intrinsically wrong, would not resolve the issue brought forth by the customer.

*4.6.1 Auto-TSG Types.* Typically, each Auto-TSG comes in with a specific TSG Type (as shown in *snippet 1*). This type can be:

1. **Informational.** This type is reserved for Auto-TSGs that retrieve general, high-level information about a specific base context (for example, a database). Those TSGs are typically important to understand the broader situation. For instance, a TSG could retrieve the history of past tickets that the customer has had with the service (or other related services), list multiple important metrics about their resource usage, and print out a list of contact information if required.

2. **Warning**. This type is typically used to display information of concern that might not fully cause unavailability or reliability problems. For instance, storage access throttling can slow down data operations without causing downtime if they are automatically retried. For some workloads, such slowdown is not a problem, especially for online analytical workloads (OLAP), which typically are more resource-intensive than transactional workloads. *Snippet 1* above was a warning-level Auto-TSG.

3. **Critical**. This type is used for Auto-TSGs that typically find an impactful problem for a given base context. For instance, repeated process crashes can impact connectivity and cause downtime to be very visible to the customer.

ARCAS will prioritize actions and information retrieved by critical TSGs, but often display the logs produced by informational and warning-level TSGs such that an on-call engineer can have a holistic picture of the state of a resource.

*4.6.2 Topical Prioritization.* However, this is not always sufficient. In many cases, the data returned by Auto-TSGs can be about inter-related problems.

For instance, backups might be missing because the database instance suffers from high storage throttling, preventing data copies from being taken. This, in turn, can cause a workload slowdown that increases the number of queued requests. As queued requests build up, more connections are created and kept open with the system, and eventually new connections get automatically rejected. Such buildup can cause a significant increase in resource consumption and cause the system to run out of memory due to various defects, causing a process to crash, and thus causing additional unavailability under the form of connectivity failures. Subject matter experts might have developed Auto-TSGs that are specific to their domain of expertise but might not fully comprehend how their piece of logic relates to others. In the example above, an Auto-TSG could track missing backups and attempt to figure out why (timeouts). Another Auto-TSG could track workload slowdown, by comparing typical request latency over time. Another one might track connection limit hits (which could typically be caused by user behavior and not a problem of the system per-se). Finally, one more could track process crashes (and what instructions caused them to crash), and another one identifies storage-related issues (in this case, high storage access throttling).

Arcas uses a large language model (LLM) to understand how these Auto-TSGs might relate to each other. For each Auto-TSG, a set of topics are extracted (for instance: connectivity, backup, performance, etc.). Those topics are used to tag Auto-TSG outputs along with their findings. Finally, an in-depth understanding of product architecture is put together by the site reliability engineer maintaining the ARCAS deployment for a given service. Those bits of information are used to form a prompt asking to sort Auto-TSGs based on their topic of interest, findings, and the global understanding of product architecture. *Snippet 2* shows a simplified version of such prompt for a fictional database system and its output using one-shot prompting. In real life, the product description would show additional internal information so that the Auto-TSG findings can be better related together.

```
INPUT:
You are given:
1. A description of a product.
2. A statement about a specific problem.
3. A list of findings about the impacted resource.

TASK:
To return a sorted list of findings given in (3) that
applies to the problem in (2) based on the product (in
(1)). Think of how findings relate to one another.

RULES:
- You can ONLY use the given findings.
- ALL findings need to be returned.
- Answer ONLY as a list in CSV format.
  - One row per finding.
  - Each row has:
    (a) the name of the finding.
    (c) a probability estimate that it applies.
    (d) a one-line explanation of why it applies.

EXAMPLE:

Product: "MageDB is an analytical database system that
stores data in memory on multiple nodes. Nodes are
provisioned for a specific database and billed to the
customer. Nodes are deployed on Kubernetes and managed by
a central control service (CCS). Customers write queries,
send them to a frontend endpoint on a single node using a
tabular data connection. An interpreter compiles and runs
queries by sending generated scala code to each backend
node. The product is updated once a week. An Upgrade
causes downtime. The data is backed up in remote storage,
and when a new node is provisioned for a database, the
in-memory data is hydrated from it."

Problem Statement: "The customer is complaining that they
receive an error 'the interpreter cannot be
instantiated'. This is new and did not occur last week
when submitting the same queries."

Auto-TSG Findings:
- Name: QueryFailures
  Category: Availability
  Explanation: Query <cd39> failed with error
  'connection lost' after running for 2:38 minutes. All
  nodes returned data but the frontend (node 541)
  failed to merge and sort the results.
  Query <sd31> failed to compile after 0.1 seconds.
- Name: ProcessCrash
  Category: Availability
  Explanation: The frontend process on node 541 crashes
  with an out-of-memory error trying to allocate space
  during a data merge operation.
- Name: ConnectivityFailure
  Category: Availability
  Explanation: we detected 2352 failed login attempts.
  All login attempts failed with 'Frontend not found.'
- Name: LongTransaction
  Category: Performance
  Explanation: session <26cd> is running a long
  transaction filing up in-memory log space.
  The transaction has been running for 4.53 hours,
  execute 26 queries and moved 10.5 GB of data.

Output:
LongTransaction, 90%, transaction taking memory
ProcessCrash, 50%, lack of memory caused frontend crash
ConnevityFailure, 20%, crash caused login failures
QueryFailure, 20%, crash causes query failures
```

*Snippet 2 Prompt example for Auto-TSG output ranking.*

ARCAS can parse the output and prioritize the findings and actions linked to each TSGs accordingly. In this example, the issue is due to a long transaction taking memory space and causing allocation failure on the frontend node. This, in turn, causes all kinds of side effects. The mitigation (which could be to kill the long-running transaction) is prioritize over others, which are then ignored. The Auto-TSG finding the crash due to out-of-memory error might propose adding more resources to the system, for instance, but such action is not correct since ARCAS was able to determine that a long-running transaction was using a lot of memory. The confidence values are used by ARCAS to determine what information to display and to know with what confidence actions can be taken. Under some threshold, actions are skipped if conflicting with each other. ARCAS can then decide to propose the actions to the SRE, effectively relegating itself as a copilot.

When the applicability of some findings is extremely small given the problem statement, ARCAS can also decide not to display some insights at all.

*4.6.3 Information Summaries.* We use the LLM for one additional pass that summarizes the prioritized findings and what action was taken, such that SREs can refer to the summary to get an overview of the issue at hand and of the findings, before referring to their details. We typically ask the LLM to split the summary into three parts:

1. The problem statement, which is not always clearly stated by customers, and for long-running incidents might have become hard to comprehend.

2. The findings and their relationship, such that a non-expert SRE can understand how those findings relate to each other.

3. Suggested actions as well as action that were taken automatically to mitigate or attempt to mitigate the incident. This goes hand in hand with the idea of prioritizing findings but separating them allows for better readability.

*Snippet 3* shows a summary produced by our LLM based on the simplified findings and example of *snippet 2*.

```
Problem Description
The MageDB analytical database system gives the customer
an error saying "the interpreter cannot be instantiated."
This is a new problem.

Findings
- A long transaction has been detected, which is taking
  up memory and is the root cause of the issue.
- A process crash (frontend) has been detected, which has
  been caused by not having enough memory.
- A connectivity failure has been detected, which is
  likely a consequence of the crash.
- Query failures have been detected, which are likely a
  consequence of the crash.

Automatic Actions
- The long-running transaction has been killed.

Suggested Actions
- Request the customer to rework the workload to reduce
  memory consumption.
- Consider upgrading the system with more memory to
  prevent future process crashes.
- Monitor the system to ensure the problem is resolved.
```
*Snippet 3 Example of Auto-TSG summary.*

## 4.7 Scheduled Auto-TSGs

So far, we have covered how we use Arcas to retrieve data for incoming incidents, to display customer-facing insights, on-demand diagnostics for the on-call team, and so on. In addition, we have covered how the finding prioritization is done under-the-hood by an LLM using a global understanding of the product architecture. However, this is not sufficient to make ARCAS a one-for-all tool for our on-call toolkit.

Using a slightly enhanced syntax, an Auto-TSG can also define a scheduled trigger. Such trigger runs at regular intervals and executes one or more telemetry queries to find if patterns are present in telemetry that warrant an in-production action, or the creation of an internal incident that needs to be looked at by an SRE. The base context for such scheduled TSG is only composed of specific timestamps, which can be used in the trigger query to make sure the data it gathers is deterministic. *Snippet 4* shows a trigger that finds databases that have long-running upgrades that do not complete, like *snippet 1*.

```yaml
Triggers:
  - Audiences:
    - Schedule
    ScheduleSettings:
      Frequency: 00:20:00
    Queries:
    - Source: Kusto
      QueryText: |
        ManagementOperations
        | where OperationName == "Upgrade"
        | where ServerName == "{ServerName}"
        | where DatabaseName == "{DatabaseName}"
        | where TimeStamp >= datetime({StartTime})
        | where TimeStamp <= datetime({TimeStamp})
        | summarize UpgradeStart = min(TimeStamp),
                    UpgradeEnd   = max(TimeStamp),
                    State = arg_max(TimeStamp, State)
                 by OperationId,
                    ServerName,
                    DatabaseName
        | extends Duration = UpgradeEnd - UpgradeStart
        | where Duration > 1h
      AddedContext:
        ServerName: string
        DatabaseName: string
        OperationId: long
        Duration: timespan
        State: string
    ScopingContext:
    - ServerName
    - DatabaseName
    NextSteps:
        - create-incident
#########################################################
Actions:
  - Name: create-incident
    Action: CreateIncident
    ThrottlingSettings:
      TimeToLive: 04:00:00
    Title: Long upgrade(s) detected for {ServerName}
    OwningService: Some Service
    OwningTeam: Release Management
  NextSteps:
   - write-resource-name
   - list-upgrade
#########################################################
Explanations:
  - Name: write-resource-name
    Explanation: >-
      This alert is for {ServerName} / {DatabaseName}
  - Name: list-upgrade
    Explanation: >-
      We detected a long upgrade `{OperationId}`.
      The current state is: **{State}**.
      The operation has been ongoing for {Duration}.
```
*Snippet 4 Example of schedule Auto-TSG.*

In this example, a query is executed every 20 minutes to check for long-running upgrades that did not finish. If a specific resource (here a database) has at least one such operation running, an incident will be created, and kept alive for at most 4 hours after the operation is finally completed. To do this, the Auto-TSG implements the concept of `ScopingKeys`, which allows to split the telemetry data returned by the trigger query into multiple logical ensembles. For each of those ensembles, an incident will be created. Explanation checks are then like *snippet 1*. The impacted resource will be written once in the created incident, and each long-running operation will be listed.

In addition, the system prevents duplicate incidents by checking if one was not already created for a specific set of scoping keys in the past. This is implemented using a moving window that is the same length as the incident time-to-live. It is thus important that the trigger frequency be smaller than the time-to-live (for our case, we enforce that the frequency is at least three times higher).

When an upgrade operation is completed, whether an SRE acted on the created incident or the situation resolved on its own, the incident will be automatically mitigated after the time-to-live as the triggering query will not find matching data anymore.

Note that this shows a simple operation that consists of creating a new incident that an SRE can then review. Other such monitors might not create an incident, but instead, call a different kind of action that can for instance act on the resource to finalize the upgrade operation. Such an operation can be to reboot processes, apply feature switches, automatically upgrade, or downgrade packages, or cancel operations altogether. These actions are queued and executed by ARCAS using an operation execution queue, which ensures that:

1. Impactful operations are throttled sufficiently to avoid large product disruption (for instance, it would be bad to have two or more such scheduled Auto-TSGs all killing the same resource at brief time intervals).
2. Conflicting operations are resolved based on detection precedence (for example, rebooting a process tied up to a resource is not needed if another operation was taken that moved the resource to an entire new set of nodes, thus causing failovers and reboots).

These operations are logged into ARCAS' telemetry so that they can then be analyzed and correlated (and are thus available for Auto-TSGs as well). *Snippet 5* shows a step that attempts to cancel the long upgrade operation, which could have been taken by the scheduled Auto-TSG proposed above either instead of or alongside incident creation.

```
Actions:
- Name: cancel-upgrade
  Action: CancelManagementOperation
  OperationId: {OperationId}
  Reason: >-
    Cancelling the long running operation {OperationId}
    for {ServerName}/{DatabaseName}, as taking too long
    to finish. The database should automatically come
    back online on the previous version.
```

*Snippet 5 Automated Action cancelling an operation.*

## 4.8 Implementation Considerations

As ARCAS needs to integrate fully with various support tools, such as internal incident management systems, user-facing insight interfaces on Azure and Fabric (for instance Azure Advisor **[13]**), as well as production interfaces to apply mitigations, the core components of ARCAS (the Auto-TSG base context extractor and the coordinators and executors) are encapsulated behind a REST API. The REST API can be called by a constellation of support tools using integrated agents. Moreover, we have designed our REST API in such a way that it allows us to interactively test and develop Auto-TSGs using a web-based enhanced YAML editor, thus allowing a subject matter expert to get a feel of the behavior of a new diagnostic in a brief time. The REST API returns a JSON structure based on a contract that can be understood by the interfacing agent of every calling tool.

*Picture 2* shows an example of execution flow against ARCAS using an interface called Incident Platform Interface, which allows to act upon an incoming incident. The incident might be customer-reported or automatically created by a legacy monitoring platform separated from ARCAS. For simplicity, this one only shows actions taken on the incident itself. However, automated mitigation actions can also be executed by this flow if required.

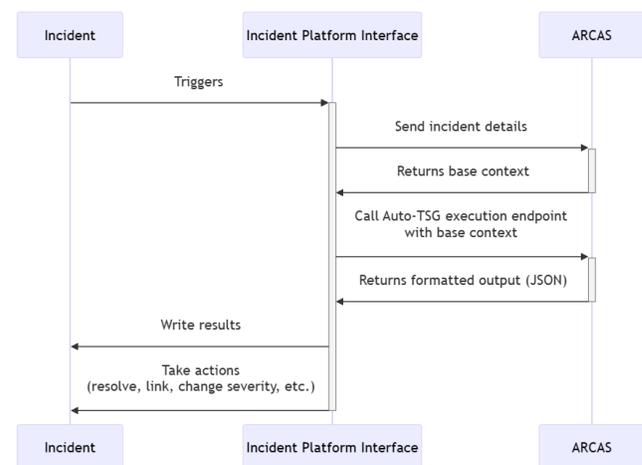

*Figure 3 Example of ARCAS API Integration.*

Note that the execution is done in two API calls. The first call uses the incident details to find the appropriate base context relating to the resource the incident is about. The second call kicks off the execution of the Auto-TSGs.

The reason behind such a split is that the Incident Platform Interface can modify the base context by injecting additional information that then becomes available to Auto-TSGs during execution. This can for instance inject a whole Auto-TSG to execute. That pattern is extensively used in case of interactive development by a subject matter expert using the web-based editor.

ARCAS handles the basic context extraction by itself to make sure that it gets or finds all the essential details needed. Some of the base contexts may not be in the incident itself, but ARCAS can use product telemetry to find them, which gives a more reliable execution starting point for Auto-TSGs. Setting up such base context extraction patterns is one of the steps needed to add a new service to ARCAS (along with, among others, telemetry accesses).

*Figure 4* shows another execution flow, in this case for scheduled TSGs, where ARCAS communicates with both the Incident Platform Interface and the Production Interface to either create incidents or apply proactive mitigations. This is the only audience for which ARCAS does not execute solely through a REST API.

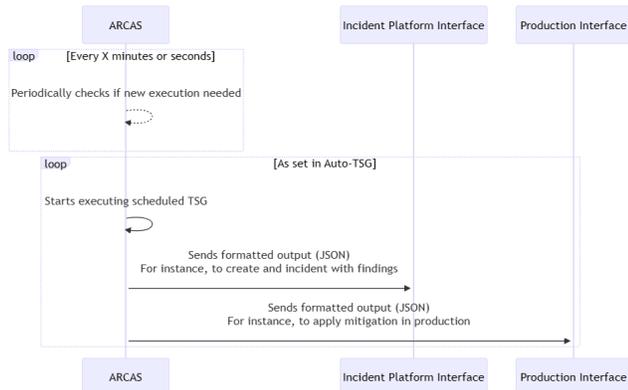

*Figure 4 Example of ARCAS Scheduled Auto-TSG Integration.*

### 4.9 Rendering

Rendering the Auto-TSG output depends on the audience and the target platform. The diagnostic data is returned as a JSON object from the REST API and processed by the calling agent.

On our troubleshooting dashboard, Auto-TSG diagnostics can be grouped by various domains of expertise. Links among diagnostics (either generated from our LLM summary or explicitly written in the Auto-TSG) allow for easier navigation. On the other hand, on support tickets, the data tends to be displayed as a list of insights ranked based on the Auto-TSG type that sourced the insight and the prioritization decided by ARCAS. Auto-TSGs with low confidence of applicability can also be hidden away. *Figure 5* shows an example of such an insight for an Asure Synapse Analytic Dedicated SQL Pool. In this case, the insight is about data volume skew among distributed data shards, which can impact performance and is against our advertised best practices **[14]**.

### 4.10 Quality Control

ARCAS is only as good as the quality of the Auto-TSGs written down by subject matter experts. To ensure a minimum quality bar, we have three main mechanisms in place.

*4.10.1 Feedback Ranking.* We gather feedback from engineers using a simple thumb-up and thumbs-down system (along with optional text feedback). That data is then shared with Auto-TSG authors. Ratings are linked not only with their Auto-TSGs, but also with the version of those TSGs. This lets us track rating changes as an auto-TSG is updated. Moreover, if the approval rate of an Auto-TSG drops below some level, we can automatically disable them and assign work items for their owners.

*4.10.2 Incident Creation Throttling.* ARCAS comes with de-duplication of incidents built-in thanks to the concept of `ScopingKeys` introduced in *section 4.7*. However, if this one is not used adequately, it can lead to noisy incident creation. To lower the incident volume, each Auto-TSG is given a default quota of incidents it can create during a period before a back-off mechanism is triggered. When such mechanism triggers, an outage tracking incident is created mentioning the spike of detection. This is useful both in case of misfires from a badly written Auto-TSG as well as when a widespread number of resources are impacted.

*4.10.3 LLM-Based Erasure.* Auto-TSGs that do not return any relevant information for a given situation can be automatically hidden by the ranking provided by our LLM-based ranking mechanism, as discussed in *section 4.6*.

In addition, Auto-TSGs are reviewed by a dedicated group of SREs before they are allowed to be deployed on our production environment to ensure that they meet a minimum quality bar and that a diagnostic creation volume assessment has been completed.

## 5 RESULTS

### 5.1 Adoption at Azure Synapse Analytics and Fabric Synapse Data Warehouse

Since the adoption of ARCAS by the Azure Synapse team (specifically for Synapse SQL Dedicated Pools and Synapse SQL Serverless), and more recently by the Microsoft Fabric team (for Synapse Lakehouse and Warehouse artifacts), multiple Auto-TSGs have been created, and the system has been widely adopted.

*Table 1* below summarizes the number of steps that have been put together among all our diagnostics as of 2023.

| Category | Count |
|---|---|
| Telemetry Triggers | > 200 |
| Telemetry Checks | > 300 |
| Explanation Blocks | > 250 |
| Automated Actions | > 150 |
| **Total Steps** | **> 1,000** |

*Table 1 Auto-TSG metrics for Azure Synapse Analytics.*

### 5.2 Estimated Time Saved

Over one month, we have thousands of API calls to execute Auto-TSGs tens of thousands of times, with the following breakdown:

| Category | API Calls | Auto-TSG Executions |
|---|---|---|
| Support (customer-facing and support engineers) | 80 % | 73 % |
| On-Demand (Web Interface) | 8 % | 11 % |
| Internal Support Tickets | 12 % | 16 % |
| **Total API Calls** | **Thousands** | **Tens of thousands** |

*Table 2 API Calls for Azure Synapse Analytics.*

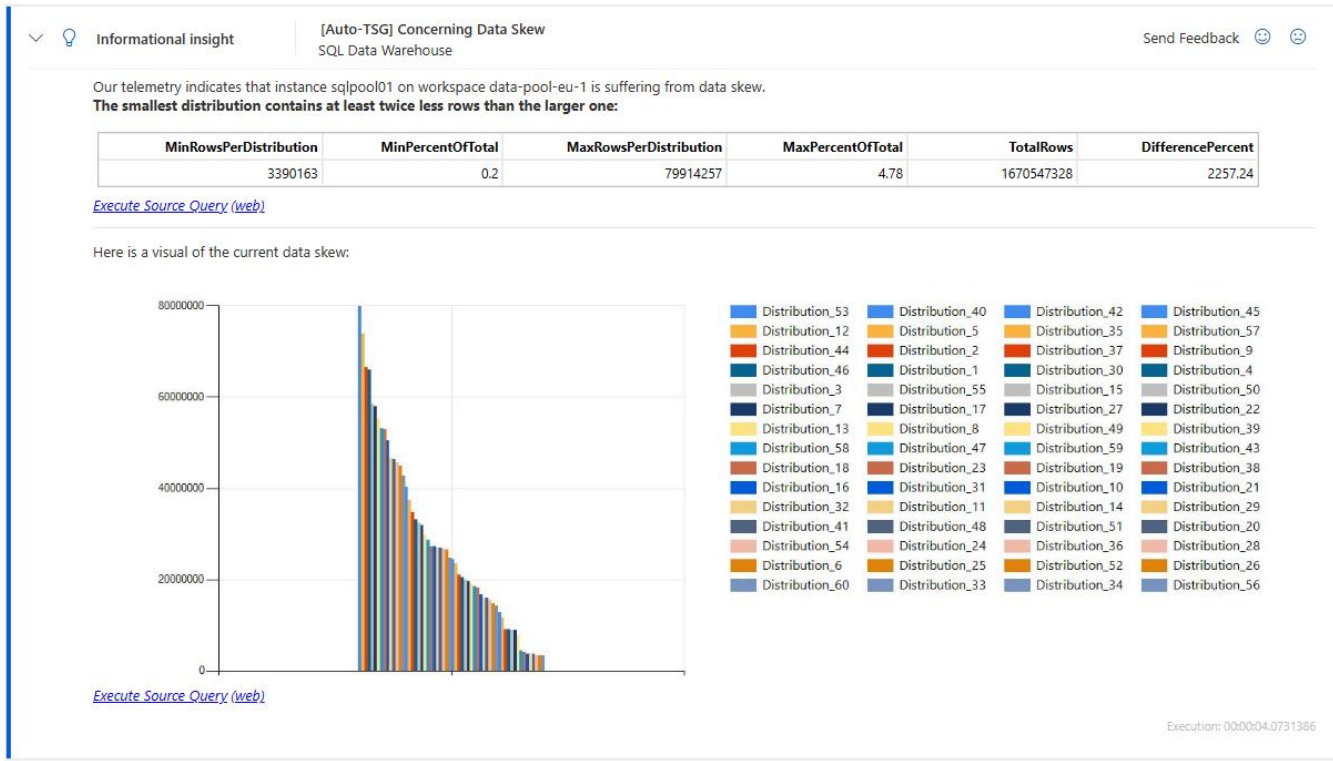

*Figure 5 Example of diagnostic for an Azure Synapse Analytic Dedicated SQL Pool on a performance-related support ticket.*

We cannot easily calculate the impact of customer-facing insights. However, we can track the time we save for SREs when they resolve incidents. This is time that benefits both customers (as we can fix incidents quicker) and our engineers, who can then focus on other tasks. Based on a 40-hour work week, we can estimate that ARCAS saves us the work equivalent of about 10 to 15 engineers for Azure Synapse and Microsoft Fabric Data Warehouse alone.

### 5.3 Auto Detected and Resolved Issues

Over one month, we have created thousands of alerts or actions using ARCAS. The exact impact on our day-to-day operation is hard to quantify, as proactive monitoring typically adds work to engineers (as opposed to proactive mitigation). However, a substantial portion of those alerts have allowed us to take proactive measures (automated or not) to mitigate issues before they become real problems or are found and reported by customers. This includes dozens of proactive detections of systemic issues during product deployments and hundreds of proactive mitigations for customer instances.

### 5.4 Engineering Adoption

Another key metric is adoption among SREs and core engineers. Since we introduced our framework, we have seen an average month-over-month growth of 16% in distinct contributors and a 19% growth in contributions. This translates into a year-over-year growth of 500% of unique contributors and a 460% increase in contributions. *Figure 6* illustrates this growth.

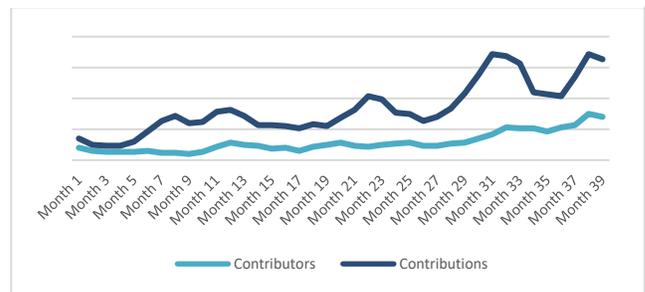

*Figure 6 ARCAS Contribution Growth.*

### 5.5 Success Stories

This section covers some anecdotal success stories using ARCAS.

*5.5.1 Login Failures.* An issue in our orchestration platform caused login failures. The issue required a complex mitigation. Using ARCAS, we were able to proactively detect those situations and route any incoming tickets directly to a responsible SRE queue, allowing for faster mitigation of hundreds of cases.

*5.5.2 Intermittent Product Failures.* Due to a new code rollout, there was a set of intermittent product failures. We stopped the rollout and were able to proactively provide a root cause summary to impacted customers using ARCAS.

*5.5.3 Microsoft Fabric Synapse Data Warehouse Availability Monitoring*. Prior to general availability, we were able to monitor

M. Demarne et al.

deployments of Microsoft Fabric Synapse Data Warehouse for potential issues and drive product reliability higher.

## 6 RELATED WORK

The landscape of incident management and monitoring tools is varied, with several platforms offering a range of capabilities that cater to the needs of complex cloud-based systems. Notable among these are Datadog **[3]** and New Relic **[4]**, which provide robust monitoring and analytics solutions, focusing on real-time data collection, visualization, and alerting. These tools, along with others such as Grafana **[10]** for dashboards and Prometheus **[9]** for monitoring and alerting, have become known in the industry for their powerful data visualization and alerting features. However, they typically require manual intervention for mitigation and do not unify data rendering alongside complex monitoring into a single development experience.

In contrast, ARCAS introduces a domain-specific language that encapsulates the expertise of subject matter experts into Auto-TSGs, which is a departure from the more general-purpose querying provided by tools like Splunk **[15]** (where the Search Processing language, SPL, resembles the Kusto Query Language **[12]**) or Elastic's Kibana **[16]**. While Splunk is renowned for its ability to ingest and analyze large volumes of data, and Kibana is regarded for its visualization capabilities on top of Elasticsearch, neither is designed with a focus on automated mitigation like ARCAS. Moreover, platforms like Nagios **[17]** and Zabbix **[18]**, which are also key players in the monitoring domain, especially around networking, offer alerting capabilities but do not inherently integrate the breadth of automated troubleshooting and mitigation that ARCAS can provides.

Automatic troubleshooting guide generation has been an area of interest in the industry. M. Shetty et al. investigate how to use Siamese networks to automatically convert manual TSGs with high precision **[19]**. We intend to take inspiration from them to automate this process.

LLM are also starting to be used more in the industry. For instance, Wang et al. introduced RCAgent **[20]**, a novel framework leveraging autonomous agents and tool-augmented large language models for efficient root cause analysis in cloud systems. Their work underscores the potential of integrating LLMs with tool capabilities to enhance decision-making and interaction within complex computing environments. The study "Recommending Root-Cause and Mitigation Steps for Cloud Incidents using Large Language Models" **[21]** builds on this momentum, showcasing the practical applications of LLMs like GPT-3.x in the context of incident management for cloud services. Through a comprehensive evaluation, the research demonstrates how fine-tuned LLMs can generate actionable recommendations for both root causes and mitigation steps. This does not, however, aim at integrating automated log discovery.

## 7 FUTURE WORK

Going forward, we plan to evolve ARCAS into a more streamlined service to provide automated onboarding for new products, thus systematizing the development of base context configuration and telemetry sources, as well as interoperability with other interfaces like the incident platform interface and the production interface mentioned in this paper. One more addition would be to have a UI-based editor that natively represents the DSL as a graph for even faster Auto-TSG development. Finally, we plan to instruct or fine-tune various LLM models to enhance our automated insight merging capabilities. Having a copilot helping subject matter experts to write Auto-TSGs is also in the work.

## 8 CONCLUSION

In this paper, we introduced ARCAS (our Automated Root Cause Analysis System), an innovative diagnostic platform designed to address the complexities of modern cloud-based data systems. ARCAS distinguishes itself by providing a comprehensive solution that integrates automated diagnostics, data rendering, monitoring, and automated mitigation within a single cohesive framework. This unified approach is facilitated by a Domain Specific Language and the use of generative AI that enables subject matter experts to encode their expertise without the need to fully comprehend the intricacies of other diagnostics or the broader system, thereby reducing the learning curve and accelerating implementation. By doing so, it streamlines the incident management process, notably reduces time-to-mitigate, and alleviates the burden on engineers. The successful deployment of ARCAS in Azure Synapse Analytics and Microsoft Fabric Synapse Data Warehouse highlights its potential as a centrally unified platform for anything related to live site.

## ACKNOWLEDGEMENTS

We are grateful to the Azure Synapse and Microsoft Fabric Synapse Data Warehouse engineering teams for their early input on ARCAS. Their feedback helped us create a system that solves many problems of traditional site reliability engineering tools. We also thank our leadership team for supporting the project when it was experimental. Lastly, we appreciate the hard work of every product and support engineer who made Azure Synapse Analytics and Microsoft Fabric Synapse Data Warehouse leaders in cloud-based data warehousing.

# REFERENCES

[1] J. Chen, X. He, Q. Lin, Y. Xu, H. Zhang, D. Hao, F. Gao, Z. Xu, Y. Dang, and D. Zhang, 2019. "An empirical investigation of incident triage for online service systems" in International Conference on Software Engineering: Software Engineering in Practice, , pp. 111–120.
[2] C. L. Forgy, 1982. "Rete: A fast algorithm for the many pattern/many objects pattern match problem".
[3] Datadog, accessed Jan 2024, available at https://www.datadoghq.com
[4] New Relic, accessed Jan 2024, available at https://newrelic.com
[5] Azure Synapse Analytics, accessed Jan 2024, available at https://azure.microsoft.com/en-us/services/synapse-analytics
[6] Microsoft Fabric, accessed Jan 2024, available at https://www.microsoft.com/en-us/microsoft-fabric
[7] React, accessed Jan 2024, available at https://react.dev
[8] Node.JS, accessed Jan 2024, available at https://nodejs.org
[9] Prometheus, accessed Jan 2024, available at https://prometheus.io
[10] Grafana, accessed Jan 2024, available at https://grafana.com
[11] YAML, YAML Ain't Markup Language, accessed Jan 2024, available at https://yaml.org
[12] Kusto Query Language, Azure Data Explorer, accessed Feb 2024, available at https://learn.microsoft.com/en-us/azure/data-explorer/kusto/query/
[13] Azure Advisor, accessed Dec 2023, available at https://learn.microsoft.com/en-us/azure/advisor/advisor-overview
[14] Guidance for designing distributed tables using dedicated SQL pool in Azure Synapse Analytics, accessed Jan 2024, available at https://learn.microsoft.com/en-us/azure/synapse-analytics/sql-data-warehouse/sql-data-warehouse-tables-distribute
[15] Splunk, accessed Jan 2024, available at https://www.splunk.com
[16] Kibana, Elastic, accessed Feb 2024, available at https://www.elastic.co/kibana
[17] Nagios, accessed Jan 2024, available at https://www.nagios.org
[18] Zabbix, accessed Jan 2024, available at https://www.zabbix.com
[19] M. Shetty, C. Bansal, S. Pramod Upadhyayula, A. Radhakrishna, and A. Gupta. 2022. "AutoTSG: Learning and Synthesis for Incident Troubleshooting" in ESEC/FSE '22, https://doi.org/10.1145/3540250.3558958
[20] Z. Wang, Z. Liu, Y. Zhang, A. Zhong, L. Fan, L. Wu, Q. Wen, 2023. „RCAgent - Cloud Root Cause Analysis by Autonomous Agents with Tool-Augmented Large Language Models". arXiv:2310.16340v1. https://arxiv.org/abs/2310.16340
[21] T. Ahmed, S. Ghosh, C. Bansal, T. Zimmermann, X. Zhang, and S. Rajmohan, "Recommending root-cause and mitigation steps for cloud incidents using large language models," in International Conference on Software Engineering, 2023, p. 1737–1749.